\documentclass[reqno,12pt]{amsart}
\usepackage{latexsym}
\usepackage{amssymb}
\usepackage{amsmath}
\usepackage{amsthm}
\usepackage{url}
\usepackage{hyperref}
\usepackage{graphicx}
\newtheorem{theorem}{Theorem}[section]
\newtheorem{lemma}[theorem]{Lemma}

\newtheorem{definition}[theorem]{Definition}
\newtheorem{remark}[theorem]{Remark}

\title{Estimating small probabilities for Langevin dynamics} 
\author{David Aristoff} 
\date{April 2012}
\address{
Department of Mathematics\\
University of Minnesota\\
Minneapolis, MN 55455}
\email{daristof@umn.edu} 

\begin{document}
\begin{abstract}
The problem of estimating small transition probabilities for overdamped Langevin dynamics is considered. 
A simplification of Girsanov's formula is obtained in which the relationship between the infinitesimal generator 
of the underlying diffusion and the change of probability measure corresponding to a change in the potential energy is 
made explicit. From this formula an asymptotic expression for transition probability densities is derived. 
Separately the problem of estimating the probability that a small noise Langevin process excapes a potential well is discussed. 
\end{abstract}

   \maketitle

\section{Introduction}\label{intro}

Let $X_t$ be a stochastic process in ${\mathbb R}^d$ satisfying the 
stochastic differential equation 
\begin{equation}\label{lang}
 dX_t = -\nabla V(X_t)\,dt + \sqrt{2\beta^{-1}}\,dW_t
\end{equation}
This is the {\it overdamped Langevin equation}. Formally, $X_t$ is a time homogeneous 
It\={o} process \cite{Oksendal} with conservative drift and constant diffusion. Intuitively, $X_t$  
represents the dynamics of large particles interacting through the potential energy $V$, 
with additional ``random'' motion driven by collisions with many small particles.
The overdamped Langevin equation can be viewed as a simplification of the well-known 
(second order) Langevin equation, which models the dynamics of a 
system of particles in contact with a heat bath at positive temperature ${\mathcal T} = (k_B\beta)^{-1}$. 
The overdamped version is obtained from a scaling limit of the Langevin equation 
in which a damping constant tends to infinity \cite{Lelievre}, \cite{Nelson}. The overdamped 
Langevin equation can then be viewed as approximating the high friction limit 
of the Langevin equation, in which no acceleration takes place. In this paper small transition 
probabilities on the process \eqref{lang} are considered. 

A useful estimate of a small probability should have an error which is much 
smaller than the probability itself. Unfortunately, standard Monte Carlo 
sampling techniques are often not useful in this sense. 
This is because for a fixed number of samples, as the probability 
$p$ being estimated approaches zero, the variance of the standard Monte Carlo 
estimate of $p$ is nearly proportional to $p$. The error, 
represented by the standard deviation, is then nearly proportional to $\sqrt p >> p$. 

Small probabilities of the process \eqref{lang} have been studied in the large $\beta$ limit 
in the context of Freidlin-Wentzell theory \cite{Freidlin}. In particular, 
the asymptotic behavior of probabilities as $\beta \to \infty$ satisfy a large 
deviations principle (LDP) \cite{Ofer}. Though the LDP by itself says nothing 
about probabilities at a fixed $\beta$, the Freidlin-Wentzell theory has 
recently been used in conjunction with optimal control theory to construct
Monte Carlo importance sampling schemes that are asymptotically optimal (as $\beta \to \infty$) 
in various senses \cite{Dupuis1}, \cite{Dupuis}, \cite{Weare}. Such schemes reduce the variance 
of standard Monte Carlo estimates by sampling with a measure under which the relevant 
event is more probable; samples are then multiplied by an appropriate factor depending 
on this measure. In general asymptotically optimal schemes of this sort 
are adaptive, with an evolving change of measure requiring significant computation 
at each time step. By contrast, non-adaptive schemes, for which the change of 
measure is fixed and impact on computation time is negligible, generally 
are not asymptotically optimal (see, however, \cite{Guasoni}). 

Introduced below is a non-adaptive importance sampling 
scheme for estimating the probability that a Langevin process escapes a potential 
well in the large $\beta$ regime. Though the analysis here is restriced to the overdamped 
case \eqref{lang}, the  
scheme can equally be used with the second order Langevin equation. 
It is shown to be asymptotically optimal in certain cases, and 
to exhibit very good (if not optimal) performance more generally. 
Estimates on its effectiveness at finite $\beta$ and asymptotically 
as $\beta \to \infty$ are given. Separately, an 
asymptotic expansion for transition probability densities 
as $t\to 0$ is proved. 

The organization of the paper is as follows. 
Background and notation are discussed and a change in measure formula 
is proved in Section~\ref{bgrd} below. 
In Section~\ref{asymp} an asymptotic expression for 
transition probabilities is proved. In Section~\ref{imp} importance 
sampling and 
the problem of estimating the probability that the process \eqref{lang} 
has exited a potential well are discussed. 
In Section~\ref{ex} a one-dimensional numerical example is provided.

\section{Background, notation and change of measure}\label{bgrd}

Here the well-known relationship between stochastic 
differential equations (SDEs) and partial differential equations (PDEs) is briefly reviewed. 
The discussion here is focused on the 
Langevin SDE
\begin{equation}\label{lang2}
 dX_t = -\nabla V(X_t)\,dt + \sqrt{2\beta^{-1}} dW_t
\end{equation}
Here $W_t$ is a $d$-dimensional Wiener process, and $V:{\mathbb R}^d \to {\mathbb R}$ 
is called the {\it potential}. Throughout it is assumed that $V \in C_b^2({\mathbb R}^d)$; 
that is, $V$ is bounded together with its (continuous) first and second order partial derivatives.
Under these conditions \eqref{lang2} has unique strong solutions 
for every initial condition as well as transition probability 
densities \cite{Koralov}.

The Langevin SDE has {\it infinitesimal 
generator} $L_V$ defined by 
\begin{equation*}
 L_V f(x) = \lim_{t\to 0}\frac{{\mathbb E}_x[f(X_t) - f(x)]}{t}
\end{equation*}
for $f \in C_b^2({\mathbb R}^d)$. 
Here ${\mathbb E}_x$ denotes expectation with respect to the 
initial condition $X_0 = x$. From It\={o}'s lemma and the dominated convergence 
theorem, one finds that 
\begin{equation}\label{L}
L_V = -\nabla V \cdot \nabla + \beta^{-1} \Delta
\end{equation}
The operator $L_V$ is closely related to probabilities of the process \eqref{lang2}. 
In particular, let $p_{t}(x,y)$ be the probability density that $X_t = y$ given that $X_0 = x$. 
(By the Markov property of the process \eqref{lang2} this determines all the transition 
probability densities.) If the second order partial derivatives of 
$V$ are all Lipschitz continuous, then for fixed $x$, $p_t(x,y)$ satisfies the PDE
\begin{equation}\label{fp}
 \frac{\partial}{\partial t}p_{t}(x,y) = L_V^* p_{t}(x,y)
\end{equation}
This is the Fokker-Planck equation \cite{Grigoriu}. Here the operator 
\begin{equation*}
L_V^* = \nabla \cdot (\nabla V\cdot) + \beta^{-1}\Delta
\end{equation*}
is formally 
adjoint to $L_V$ and in \eqref{fp} is assumed to act only on the the $y$-component of $p_t(x,y)$. 
In principle by numerically solving the Fokker-Planck equation one obtains the transition probability 
densities, but this is impractical when the dimension $d$ is large.

Let ${\mathbb P}$ be the reference probability measure under which $X_t$ satisfies 
\begin{equation*}
 dX_t = -\nabla V(X_t)\,dt + \sqrt{2\beta^{-1}} dW_t
\end{equation*}
One might ask how the measure ${\mathbb P}$ changes if $V$ is replaced 
by another potential $\tilde V$. In general this question is answered by 
Girsanov's theorem \cite{Grigoriu}, \cite{Girsanov}. However, the special structure of 
the overdamped Langevin equation allows for a useful simplification to the well-known 
Girsanov formula. In fact in Theorem~\ref{theorem1} below it is shown that the change 
in probability measure has a simple relationship with the infinitesimal generators $L_V$ and $L_{\tilde V}$:
\vskip10pt

\begin{theorem}\label{theorem1}
Assume $V, \tilde V \in C_b^2({\mathbb R}^d)$. 
Let $\tilde {\mathbb P}$ be the reference measure under which $X_t$ 
satisfies 
\begin{equation*}
 dX_t = -\nabla {\tilde V}(X_t)\,dt + \sigma\, d{\tilde W}_t, \hskip10pt X_0 = x_0, \hskip10pt 0\le t \le T
\end{equation*}
where $\sigma = \sqrt{2\beta^{-1}}$ and ${\tilde W}_t$ is a $d$-dimensional Wiener process. 
Define $\mathbb {P}$ by 
\vskip5pt

\begin{equation}\label{chg1}
\frac{d{\mathbb P}}{d{\tilde{\mathbb P}}} = 
\exp\left[\sigma^{-2}\begin{pmatrix} V(x_0) - V(X_T) - \left[{\tilde V}(x_0) - {\tilde V}(X_T)\right] \\ 
+ \frac{1}{2}\int_0^T \left[(L_V + L_0)V(X_s) - (L_{\tilde V}+L_0){\tilde V}(X_s)\right]ds\end{pmatrix}\right]
\end{equation}
\vskip8pt

\noindent where $L_\cdot$ is given by \eqref{L}. Then under $\mathbb {P}$, $X_t$ satisfies 
\begin{equation*}
 dX_t = -\nabla {V}(X_t)\,dt + \sigma\,d{W}_t, \hskip10pt X_0 = x_0, \hskip10pt 0\le t \le T
\end{equation*}
where ${W}_t$ is a $d$-dimensional $\mathbb {P}$-Wiener process.
\end{theorem}
\vskip10pt
\begin{remark}\label{remark1}
In the above, $L_0 \equiv \beta^{-1}\Delta$.
\end{remark}
\vskip10pt

\begin{proof}
Let $U = {\tilde V} - V$, and define $\mathbb P$ by 
\begin{equation}\label{chg2}
  \frac{d{\mathbb P}}{d{\tilde{\mathbb P}}} = \exp\left[\sigma^{-1}\int_0^T \nabla U(X_s)\cdot d{\tilde W}_s - \frac{\sigma^{-2}}{2}\int_0^T \nabla U(X_s) \cdot \nabla U(X_s)\,ds\right] 
\end{equation}
By Girsanov's theorem, under $\mathbb P$ the process $X_t$ satisfies 
\begin{equation*}
 dX_t = -\nabla V(X_t)\,dt +  \sigma d{W}_t, \hskip10pt X_0 = x_0, \hskip20pt t \in [0,T]
\end{equation*}
By It\={o}'s lemma, under $\tilde {\mathbb P}$, 
\begin{align}
dU(X_s) &= \nabla U(X_s)\cdot dX_s + \frac{1}{2} dX_s'\, \nabla^2 U(X_s)\, dX_s \label{gr1} \\
        &= -\nabla U(X_s)\cdot \nabla {\tilde V}(X_s)\,ds + \sigma \nabla U(X_s)\cdot d{\tilde W}_s + \frac{\sigma^2}{2} \Delta U(X_s)\,ds \label{gr2}
\end{align}
where $\nabla^2 U$ denotes the Hessian matrix of $U$, and $dX_s'$ is the transpose of $dX_s$. 
Rearranging, multiplying by $\sigma^{-2}$, and using the integral form of \eqref{gr1}-\eqref{gr2}, this becomes 
\begin{align}
&\sigma^{-1}\int_0^T \nabla U(X_s)\cdot d{\tilde W}_s \label{simp1} \\
&= \sigma^{-2}[U(X_T) - U(x_0)] + \sigma^{-2}\int_0^T \nabla U(X_s)\cdot \nabla {\tilde V}(X_s)\,ds - \frac{1}{2}\int_0^T \Delta U(X_s)\,ds \label{simp2}
\end{align}
Substituting \eqref{simp1}-\eqref{simp2} into \eqref{chg2} and simplifying, 
\begin{align}\label{last0}
    &\frac{d{\mathbb P}}{d{\tilde{\mathbb P}}} = \exp\left[\sigma^{-1}\int_0^T \nabla U(X_s)\cdot d{\tilde W}_s - \frac{\sigma^{-2}}{2}\int_0^T \nabla U(X_s) \cdot \nabla U(X_s)\,ds\right] \\
 &= \exp\left[\sigma^{-2}\left(U(X_T)-U(x_0) + \frac{1}{2}\int_0^T 
\left(\nabla {\tilde V}\cdot \nabla {\tilde V} - \nabla {V}\cdot \nabla {V} - \sigma^2\Delta U\right)(X_s) \,ds\right) \right]  \label{last}
\end{align}
By comparing \eqref{last0}-\eqref{last} with \eqref{chg1}, the result follows.
\end{proof}
\vskip10pt

Although Girsanov's formula and It\={o}'s lemma can be used with any It\={o} process \cite{Grigoriu}, 
in the proof of Theorem~\ref{theorem1} the assumptions that the change in drift (here $\nabla {\tilde V} - \nabla V$) 
is conservative and that the diffusion matrix (here $\sigma I_d$) is a constant multiple of the 
identity matrix are essential. The result can be generalized slightly: 
\vskip10pt

 \begin{theorem}\label{theorem1b}
Assume $V \in C_b^2({\mathbb R}^d)$ and $F: {\mathbb R}^d \to {\mathbb R}^d$ 
is Lipschitz continuous. 
Let ${\mathbb P}^{ref}$ be the reference measure under which $X_t$ 
satisfies 
\begin{equation*}
 dX_t = F(X_t)\,dt + \sigma\, d{W}_t^{ref}, \hskip10pt X_0 = x_0, \hskip10pt 0\le t \le T
\end{equation*}
where ${ W}_t^{ref}$ is a $d$-dimensional Wiener process and $\sigma>0$. 
Define $\mathbb {P}$ by 
\vskip5pt

\begin{equation*}
\frac{d{\mathbb P}}{d{{\mathbb P}}^{ref}} = 
\exp\left[\sigma^{-2}\left(V(x_0) - V(X_T) + \frac{1}{2}\int_0^T \left(L_V -L_0 + 2L^{ref}\right)V(X_s)\,ds\right)\right]
\end{equation*}
\vskip5pt

\noindent where 
\begin{equation*}
 L^{ref} = F \cdot \nabla + \frac{\sigma^2}{2} \Delta
\end{equation*}
is the infinitesimal generator of the reference process. 
Then under $\mathbb {P}$, $X_t$ satisfies 
\begin{equation*}
 dX_t = -\nabla {V}(X_t)\,dt + F(X_t)\,dt + \sigma\,d{W}_t, \hskip10pt X_0 = x_0, \hskip10pt 0\le t \le T
\end{equation*}
where ${W}_t$ is a $d$-dimensional $\mathbb {P}$-Wiener process.
\end{theorem}
\vskip10pt

\noindent The proof of Theorem~\ref{theorem1b} is similar to that of Theorem~\ref{theorem1} and is 
therefore omitted. 

Note that much intuition can be gained out of a simple inspection of the formula \eqref{chg1}. 
For example if $T$, $\delta$ are small and 
\begin{equation*}
 {\mathcal A} = \{X_\cdot\,:\, X_T \in B_{\delta}(y)\}
\end{equation*}
where $B_{\delta}(y)$ is a ball of radius $\delta$ around $y$, then
\begin{equation}\label{insp}
 {\mathbb P}({\mathcal A}) \approx \exp\left[\sigma^{-2}\left(V(x_0) - V(y) - \left[{\tilde V}(x_0)-{\tilde V}(y)\right]\right)\right]{\tilde {\mathbb P}}({\mathcal A})
\end{equation}
In particular, if $\tilde V = 0$ then the probability on the right hand side 
of \eqref{insp} can be written as an integral of a Gaussian 
density; this suggests an estimate of asymptotic transition probabilities which is pursued in the next section.

\section{Asymptotic transition probabilities}\label{asymp}

Consider the transition probability density  
$p_{t}(x,y)$ of the process 
\begin{equation}\label{lange}
 dX_t = -\nabla V(X_t)\,dt + \sigma\,dW_t
\end{equation}
Recall $p_t(x,y)$ is the conditional probability density that $X_t = y$ given that 
$X_0 = x$. Notice that if $V=0$ in \eqref{lange} then $X_t = \sigma W_t$ and 
\begin{equation*}
p_t(x,y) = {(2\pi \sigma^2 (t-s))^{-d/2}}\exp\left(-\frac{|y-x|^2}{2\sigma^2 (t-s)}\right)
\end{equation*} 
In the following theorem Theorem~\ref{theorem1} 
is used to estimate transition probability densities for a generic potential.
\vskip10pt

\begin{theorem}
 \label{theorem2}
Assume $V \in C_b^2({\mathbb R}^d)$ and $(L_V+L_0) V$ is Lipchitz continuous with Lipschitz constant $K$. 
Let $p_{t}(x,y)$ be the transition probability density of the process \eqref{lange}.
Define 
\begin{equation*}
 \psi_{x,y}(r) = (1-r)x + ry
\end{equation*}
Then for any $\delta > 0$, 
\begin{align*}
&p_{t}(x,y) \ge \\
&\left(\exp\left[\sigma^{-2}\left(V(x)-V(y) + \frac{t}{2}\int_0^1 (L_V+L_0) V(\psi_{x,y}(r))\,dr - M_1 \delta t\right)\right] - M_2 \gamma(\delta,t)\right)\rho_{t}(y-x)
\end{align*}
and  
\begin{align*}
&p_{t}(x,y) \le \\
&\left(\exp\left[\sigma^{-2}\left(V(x)-V(y) + \frac{t}{2}\int_0^1 (L_V+L_0) V(\psi_{x,y}(r))\,dr + M_1 \delta t\right)\right] + M_2 \gamma(\delta,t)\right)\rho_{t}(y-x)
\end{align*}
where 
\begin{align*}
 &M_1 = \frac{1}{2}\sqrt{d} K  \\
 &M_2 = 2d\exp\left(\sigma^{-2}\left[V(x) - V(y) + \frac{t}{2} \sup |(L_V+L_0) V|\right]\right) \\
 &\gamma(\delta,t) = \exp\left(-\frac{2\delta^2}{\sigma^2 t}\right) \\
 &\rho_{t}(x) = {(2\pi\sigma^2 t)^{-d/2}}\exp\left(-\frac{|x|^2}{2\sigma^2 t}\right)
\end{align*}
In particular, as $t \to 0^+$, 
\begin{align}
&p_{t}(x,y) = \\
&\left(\exp\left[\sigma^{-2}\left(V(x)-V(y) + \frac{t}{2}\int_0^1 (L_V+L_0)V(\psi_{x,y}(r))\,dr\right)\right]+O\left(t^{\alpha+1}\right)\right)\rho_{t}(y-x) \label{correction}
\end{align}
for any $\alpha \in \left(0,\frac{1}{2}\right)$.
\end{theorem}
\vskip10pt

\begin{remark}\label{remark2}
 Note that $\rho_{t}(y-x)$ is the transition probability density of the process $\sigma W_{t}$.
\end{remark}
\vskip10pt

Theorem~\ref{theorem2} can be seen a a first-order correction to transition probability densities 
when a conservative drift is added to the process
\begin{equation}
 dX_t = \sigma \,dW_t
\end{equation}
as in \eqref{lange}. The term in parentheses 
in \eqref{correction} gives the correction corresponding to the addition of the 
drift $-\nabla V$. Note that when $t$ is small, the correction is dominated by the term 
$\exp[\sigma^{-2}(V(x)-V(y))]$, which depends only on the change in potential energy from 
$x$ to $y$. 

Using Theorem~\ref{theorem1b}, the asymptotic result of Theorem~\ref{theorem2} can be generalized as follows: 
\vskip10pt

\begin{theorem}
 \label{theorem2b}
Let $L^{ref}$ and $p_t^{ref}(x,y)$ be the infinitesimal generator and transition probability density of the reference process 
\begin{equation}
dX_t = F(X_t)\,dt + \sigma\,dW_t
\end{equation}
where $W_t$ is a $d$-dimensional Wiener process and $F:{\mathbb R}^d \to {\mathbb R}^d$ 
is Lipschitz continuous. Assume $V \in C_b^2({\mathbb R}^d)$ and $(L_V- L_0 + 2L^{ref})V$ is bounded and 
Lipchitz continuous. Let $p_{t}(x,y)$ be the transition probability density of the process
\begin{equation}
dX_t = -\nabla V(X_t)\,dt + F(X_t)\,dt + \sigma\,dW_t
\end{equation}
and define
\begin{equation*}
 \psi_{x,y}(r) = (1-r)x + ry
\end{equation*}
Then as $t \to 0^+$, 
\begin{align*}
&p_{t}(x,y) = \\
&\left(\exp\left[\sigma^{-2}\left(V(x)-V(y) + \frac{t}{2}\int_0^1 \left(L_V-L_0+2L^{ref}\right)V(\psi_{x,y}(r))\,dr\right)\right]+O\left(t^{\alpha+1}\right)\right)p_{t}^{ref}(x,y) 
\end{align*}
for any $\alpha \in \left(0,\frac{1}{2}\right)$.
\end{theorem}
\vskip10pt

The proof of Theorem~\ref{theorem2} is a consequence of Theorem~\ref{theorem1} 
and the following lemmas:
\vskip10pt

\begin{lemma}
 \label{lemma1}
Let ${\mathbb P}$ be the measure under which $X_t$ satisfies
\begin{equation*}
 dX_t = \sigma\, dW_t,\hskip10pt X_0 = x
\end{equation*}
where $W_t$ is a $d$-dimensional Wiener process. Fix $t>0$, 
$x = (x_1,x_2,...,x_d) \in {\mathbb R}^d$, $y = (y_1,y_2,...,y_d) \in {\mathbb R}^d$, and $\delta>0$. Define 
\begin{equation*}
N_{\delta,t}^{x,y}(r) = \prod_{k=1}^d \left(\left(1-\frac{r}{t}\right)x_k + \frac{r}{t}y_k -\delta, \left(1-\frac{r}{t}\right)x_k + \frac{r}{t}y_k + \delta\right)
\end{equation*}
Then 
\begin{equation*}
 {\mathbb P}\left(X_r \in N_{\delta,t}^{x,y}(r), \,\, 0\le r \le t \,\big|\, X_t = y\right) \ge 1-2d\exp\left({-\frac{2\delta^2}{\sigma^2 t}}\right)
\end{equation*}
\end{lemma}
\vskip10pt
\begin{proof}
With $X_t^k$ the $k$th component of $X_t$, a well-known formula of Siegmund (\cite{Wang}, \cite{Siegmund}) leads to 
\begin{align*}
 &{\mathbb P}\left(X_r^k < \left(1-\frac{r}{t}\right)x_k + \frac{r}{t}y_k + \delta, \,\, 0\le r \le t \,\big| \,X_t^k = y_k\right) \\
 &=1 - \exp\left({-\frac{2\delta^2}{\sigma^2 t}}\right) \\
 &={\mathbb P}\left(X_r^k > \left(1-\frac{r}{t}\right)x_k + \frac{r}{t}y_k-\delta, \,\, 0\le r \le t\, \big|\, X_t^k = y_k\right)
\end{align*}
The result follows by subadditivity.
\end{proof}
\vskip10pt

\begin{lemma}
\label{lemma2}
Assume $G:{\mathbb R}^d \to {\mathbb R}$ is Lipschitz continuous with Lipschitz constant $K_G$. Fix 
$t >0$, $x,y \in {\mathbb R}^d$ and $\delta>0$. 
If $X_r \in N_{\delta,t}^{x,y}(r)$ for all $r\in [0,t]$ then 
\begin{equation*}
 \left|\int_0^t G(X_r)\,dr - t \int_0^1 G((1-r)x + ry)\,dr\right| \le \sqrt{d}\delta K_G t
\end{equation*}
\end{lemma}
\vskip10pt

\begin{proof} 
Note that if $X_r \in N_{\delta,t}^{x,y}(r)$ for all $r\in [0,t]$, then 
\begin{eqnarray*}
 \left|\int_0^t G(X_r)\,dr - t \int_0^1 G((1-r)x+ry)\,dr\right|
&=& \left|\int_0^t G(X_r)\,dr - \int_0^t G\left(\left(1-\frac{r}{t}\right)x + \frac{ry}{t}\right)\,dr\right| \\
&\le& \int_0^t \left|G(X_r) - G\left(\left(1-\frac{r}{t}\right)x + \frac{ry}{t}\right)\right|\,dr \\
&\le& \int_0^t \sqrt{d} \delta K_G \,dr \\
&=& \sqrt{d}\delta K_G t
\end{eqnarray*}
\end{proof}
\vskip10pt

\noindent {\it Proof of Theorem~\ref{theorem2}.} 
Let $\tilde {\mathbb P}$ be the reference measure under which $X_t$ satisfies 
\begin{equation*}
 dX_t =  \sigma \,d{\tilde W}_t, \hskip10pt X_0 = x, \hskip10pt 0\le t \le T
\end{equation*}
where ${\tilde W}_t$ is a $d$-dimensional Wiener process, and 
let $\tilde {\mathbb E}$ be the corresponding expectation. 
For $y=(y_1,y_2,...,y_d) \in {\mathbb R}^d$ and $h>0$ define 
\begin{equation*}
S_{y,h} = \prod_{k=1}^d \,\left[y_k, y_k + h\right)
\end{equation*}
Using Theorem~\ref{theorem1} with ${\tilde V} = 0$ yields
\begin{align*}
 &{\mathbb P}(X_{t} \in S_{y,h}) = \\
&{\tilde {\mathbb E}}
\left[\exp\left(\sigma^{-2}\left(V(x)-V(X_{t}) + \frac{1}{2}\int_0^{t} (L_V+L_0) V(X_r)\,dr\right)\right)\,1_{\{X_{t} \in S_{y,h}\}}\right]
\end{align*}
such that under ${\mathbb P}$, $X_t$ satisfies 
\begin{equation*}
 dX_t = -\nabla V(X_t)\,dt + \sigma \,dW_t, \hskip10pt X_0 = x, \hskip10pt 0\le t \le T
\end{equation*}
with $W_t$ a $d$-dimensional ${\mathbb P}$-Wiener process. Now 
\begin{align}\label{lim1}
 &\frac{{\mathbb P}(X_{t} \in S_{y,h})}{h^d} \\
&= 
\frac{{\tilde {\mathbb E}}\left[{\exp\left({\sigma^{-2}\left(V(x)-V(X_{t}) + \frac{1}{2}\int_0^{t} (L_V+L_0) V(X_r)\,dr\right)}\,1_{\{X_{t} \in S_{y,h}\}}\right)}\right]}
{{\tilde{\mathbb P}}(X_{t} \in S_{y,h})}\cdot\frac{{\tilde{\mathbb P}}(X_{t} \in S_{y,h})}{h^d} \label{lim2}
\end{align}
Taking limits in \eqref{lim1}-\eqref{lim2} as $h\to 0$ gives 
\begin{align*}
 &p_t(x,y) = \\
&{\tilde {\mathbb E}}\left[\left.\exp\left(\sigma^{-2}\left(V(x)-V(y) + \frac{1}{2}\int_0^{t} (L_V+L_0) V(X_r)\,dr\right)\right)\,\right|\, X_{t} = y\right] \rho_{t}(y-x)
\end{align*}
The first statement of the theorem follows from Lemmas~\ref{lemma1}-\ref{lemma2} with 
$G = (L_V + L_0)V$. The last statement follows by 
taking $\delta = t^\alpha$.\qed
\vskip10pt

The proof of Theorem~\ref{theorem2b}, which is omitted, is similar to the proof of 
Theorem~\ref{theorem2} and relies on the fact that the 
exit probabilities of the pinned diffusion of Lemma~\ref{lemma1} retain the same 
asymptotics as $t\to 0$ with the addition of a drift $F$ (see Theorem 2.1 of \cite{Baldi}).

\section{Importance sampling and exiting a well}\label{imp}

Here the problem of estimating a small escape probability ${\mathbb P}({\mathcal A})$ of the 
process \eqref{lang} is considered. In standard Monte Carlo, one estimates ${\mathbb P}({\mathcal A})$ 
by taking the average number of samples, out of some total $N$, for which the event $\mathcal A$ is observed. 
More precisely, the standard Monte Carlo approximation of ${\mathbb P}({\mathcal A})$ is  
\begin{equation}\label{theta1}
 \Theta \equiv \frac{1}{N}\sum_{n=1}^N 1_{\mathcal A}^n
\end{equation}
where $1_{\mathcal A}^n$ are i.i.d. random variables with the same distribution (under $\mathbb P$) as the indicator 
function $1_{\mathcal A}$. This estimate has expected value 
\begin{equation*}
 {\mathbb E}(\Theta) = {\mathbb P}({\mathcal A})
\end{equation*}
and variance 
\begin{equation*}
 \hbox{Var}(\Theta) = \frac{\hbox{Var}(1_{\mathcal A})}{N} 
\end{equation*}
where 
 \begin{equation}\label{var1}
 \hbox{Var}(1_{\mathcal A}) = {\mathbb P}({\mathcal A}) - {\mathbb P}({\mathcal A})^2
\end{equation}
The {\it relative error} of $\Theta_N$ is its standard deviation divided its expected value:
\begin{equation*}
 \hbox{Relative Error}(\Theta) \equiv \frac{\sqrt{\hbox{Var}(\Theta)}}{{\mathbb P}({\mathcal A})} 
= \frac{1}{\sqrt{N}}\sqrt{\frac{{\mathbb P}({\mathcal A})}{{\mathbb P}({\mathcal A})^2}-1}
\end{equation*}
The relative error blows up for fixed $N$ as ${\mathbb P}({\mathcal A}) \to 0$, making 
the estimate \eqref{theta1} useless for a fixed computational effort if ${\mathbb P}({\mathcal A})$ 
is very small. 

An alternative to standard Monte Carlo sampling is {\it importance 
sampling} (see e.g. \cite{Asmussen}, \cite{Bucklew}), in which one chooses another 
probability measure $\tilde {\mathbb P}$ for sampling. One then estimates ${\mathbb P}({\mathcal A})$ 
by taking the average number of samples (out of $N$) for which ${\mathcal A}$ has occurred, 
such that each sample is weighted by the factor ${d\mathbb P}/{d\tilde {\mathbb P}}$. More 
precisely, an unbiased {\it importance sampling estimator} for ${\mathbb P}({\mathcal A})$ is 
\begin{equation}\label{theta2}
 {\tilde \Theta} \equiv \frac{1}{N}\sum_{n=1}^N \left(\frac{d\mathbb P}{d\tilde {\mathbb P}}1_{\mathcal A}^n\right)
\end{equation}
where $1_{\mathcal A}^n$ are i.i.d. random variables with the same distribution (under $\tilde{\mathbb P}$)  
as $1_{\mathcal A}$. Here $\mathbb P$ must be absolutely continuous with respect to $\tilde {\mathbb P}$. 
${\tilde \Theta}$ is called {\it unbiased} because 
\begin{equation*}
 {\mathbb P}({\mathcal A}) = {\mathbb E}\left[1_{\mathcal A}\right] 
= {\tilde {\mathbb E}}\left[\frac{d\mathbb P}{d\tilde {\mathbb P}}\,1_{\mathcal A}\right]
\end{equation*}
which implies 
\begin{equation*}
 {\mathbb P}({\mathcal A}) = {\mathbb E}\left[\Theta\right] = {\tilde {\mathbb E}}\left[{\tilde \Theta}\right]
\end{equation*}
where ${\tilde {\mathbb E}}$ is the expectation corresponding to $\tilde{\mathbb P}$. 
To optimally reduce the number of samples necessary to achieve a given error, one wants to minimize the variance
\begin{equation*}
{\tilde {\hbox{Var}}}({\tilde \Theta}) = \frac{1}{N}{\tilde{\hbox{Var}}}\left(\frac{d\mathbb P}{d\tilde {\mathbb P}}\,1_{\mathcal A}\right)
\end{equation*}
subject to constraints of feasibility. Here
\begin{equation}\label{var2}
 {\tilde{\hbox{Var}}}\left(\frac{d\mathbb P}{d\tilde {\mathbb P}}\,1_{\mathcal A}\right) = {\tilde {\mathbb E}}\left[\left(\frac{d\mathbb P}{d\tilde {\mathbb P}}\right)^2\,1_{\mathcal A}\right] - {\mathbb P}({\mathcal A})^2 
= {\mathbb E}\left[\frac{d\mathbb P}{d\tilde {\mathbb P}}\,1_{\mathcal A}\right] - {\mathbb P}({\mathcal A})^2
\end{equation}
One would hope, for instance, that the variance is greatly reduced compared with standard Monte Carlo, that is, 
\begin{equation}\label{reduct}
\frac{{\tilde {\hbox{Var}}}({\tilde \Theta})}{\hbox{Var}(\Theta)}
= \frac{{\tilde{\hbox{Var}}}\left(\frac{d\mathbb P}{d\tilde {\mathbb P}}\,1_{\mathcal A}\right)}{\hbox{Var}(1_{\mathcal A})} \le C << 1
\end{equation}
Another important quantity is the relative error 
\begin{equation}\label{relerror}
 \hbox{Relative Error}({\tilde \Theta}) \equiv 
\frac{\sqrt{{\tilde{\hbox{Var}}}({\tilde \Theta})}}{{\mathbb P}(\mathcal A)}
= \frac{1}{\sqrt{N}}\sqrt{\frac{{\mathbb E}\left(\frac{d{\mathbb P}}{d{\tilde{\mathbb P}}}\,1_{\mathcal A}\right)}{{\mathbb P}({\mathcal A})^2}-1}
\end{equation}
To minimize the relative error, one wants to minimize the quantity 
\begin{equation}\label{R}
 \Lambda \equiv \frac{{\mathbb E}\left(\frac{d{\mathbb P}}{d{\tilde{\mathbb P}}}\,1_{\mathcal A}\right)}{{\mathbb P}({\mathcal A})^2}
\end{equation}

In general it is very difficult to prove an inequality 
like \eqref{reduct}, or useful bounds on \eqref{relerror}-\eqref{R}, outside of certain asymptotic regimes. 
Examined below is the {\it small noise} regime of the overdamped Langevin equation, defined by
\begin{equation*}
 dX_t^\epsilon = -\nabla V(X_t^\epsilon)\,dt + \sqrt{\epsilon}\,dW_t
\end{equation*}
where $\epsilon$ is a small parameter. The small noise regime can be thought of as a nearly 
deterministic version of the SDE, where the dynamics are dominated by the potential energy and 
diffusive effects are small.

In the below the reduction in variance from using \eqref{theta2} instead of 
\eqref{theta1} for estimating probabilities in the small noise regime of 
the overdamped Langevin SDE 
is considered. Though the analysis is restricted to the small noise 
overdamped Langevin equation, the method itself is applicable to the 
second-order Langevin equation. The scheme involves only changes in 
measure ${\mathbb P} \to \tilde{\mathbb P}$ corresponding 
to a fixed change in the potential $V \to {\tilde V}$. 
That is, the sampling measure $\tilde{\mathbb P}$ will correspond to the process 
\begin{equation}\label{smallnoise2}
 dX_t^\epsilon = -\nabla {\tilde V}(X_t^\epsilon)\,dt + \sqrt{\epsilon}\,d{\tilde W}_t, \hskip10pt X_0 = x_0, \hskip10pt 0\le t \le T
\end{equation}
whereas the target measure ${\mathbb P}$ will correspond to 
\begin{equation}\label{smallnoise1}
 dX_t^\epsilon = -\nabla V(X_t^\epsilon)\,dt + \sqrt{\epsilon}\,dW_t, \hskip10pt X_0 = x_0, \hskip10pt 0\le t \le T
\end{equation}
Here and throughout the dependence of $\mathbb P$ and $\tilde{\mathbb P}$ on $\epsilon$ is suppressed. 
In Theorem~\ref{theorem3} it is shown that for estimating the probability 
of escaping a potential well, an exponential reduction in variance 
(compared to standard Monte Carlo) can be achieved simply by taking 
a sampling potential $\tilde V$ which reduces the depth of the well. The magnitude of the 
reduction in variance is closely related to the difference $V(x_0)-{\tilde V}(x_0)$. 
The scheme allows for the well to be ``inverted,'' and in fact in Theorem~\ref{theorem4} 
it is shown that under certain conditions this creates an {\it asymptotically optimal} 
reduction in variance, in the sense that \cite{Dupuis}
\begin{equation}\label{above}
 \lim_{\epsilon \to 0}  \epsilon \log  \Lambda = 0
\end{equation}
The limit in \eqref{above} is optimal because for any ${\mathbb P}$, ${\tilde {\mathbb P}}$, 
and ${\mathcal A}$, Jensen's inequality and \eqref{var2} imply $\Lambda \ge 1$.

Below $\Lambda_\epsilon$ is written for $\Lambda$ defined in \eqref{R}, and $\Theta_\epsilon$, 
${\tilde {\Theta}}_\epsilon$ are written for $\Theta$, $\tilde {\Theta}$ defined in \eqref{theta1}, \eqref{theta2}, 
to emphasize the dependence of these objects on $\epsilon$. Events of the following type will be considered:
\begin{definition}
 \label{definition1}
Let $D\subset {\mathbb R}^d$ be a bounded open set such that $\partial D$ is a simple closed curve,   
and define 
\begin{equation*}
 {\mathcal A}_\epsilon \equiv \{X_\cdot^\epsilon \,:\, X_T^\epsilon \notin D\}
\end{equation*}
\end{definition}
The following is the main result of this section.
\vskip10pt

\begin{theorem}
 \label{theorem3}
Assume $V \in C_b^2({\mathbb R}^d)$, ${\tilde V} \in C_b^2({\mathbb R}^d)$, such that:
\begin{align*}
 &\hbox{(i)} \,\,\,\, V(x_0) < {\tilde V}(x_0) \\
 &\hbox{(ii)} \,\,\,\, |\nabla {\tilde V}(x)| \le |\nabla V(x)| \hskip10pt \hbox{for all } x \in D \\
 &\hbox{(iii)} \,\,\,\, {\tilde V}(x) = V(x) \hskip10pt \hbox{for all } x \notin D \\
\end{align*}
and define
 \begin{equation*}
M = \frac{1}{2}\sup_{x \in D}\left(\Delta V(x)-\Delta {\tilde V}(x)\right) 
\end{equation*}
Let $\tilde {\mathbb P}$ be the reference measure under which $X_t^\epsilon$ satisfies \eqref{smallnoise2}, 
and define ${{\mathbb P}}$ as in \eqref{chg1}. \textup{(}The dependence of $\mathbb P$ and $\tilde {\mathbb P}$ 
on $\epsilon$ is suppressed.\textup{)} Then under 
${{\mathbb P}}$, $X_t^\epsilon$ satisfies \eqref{smallnoise1}. If 
\begin{equation*}
 {\tilde V}(x_0)-V(x_0) \ge \epsilon TM
\end{equation*}
then
\textup{
\begin{equation*}
 \frac{{\tilde {\hbox{Var}}}({\tilde \Theta}_\epsilon)}{\hbox{Var}(\Theta_\epsilon)}
\le e^{\epsilon^{-1}\left(V(x_0) - {\tilde V}(x_0)\right) + TM}
\end{equation*}}
Furthermore 
\begin{equation}\label{asympopt}
 \lim_{\epsilon\to 0} \epsilon \log \Lambda_\epsilon
\le V(x_0) - {\tilde V}(x_0) + I_V(x_0)
\end{equation}
where 
\begin{equation}\label{defI}
 I_V(x_0) \equiv \inf\left\{\frac{1}{2}\int_0^T \left|\dot \phi(t)+\nabla V(\phi(t))\right|^2 dt\,\,: \,\, \phi \in H_{x_0, T}^1,\, \phi(T) \notin D\right\}
\end{equation}
with 
\begin{equation*}
 H_{x_0, T}^1 \equiv \left\{\phi:[0,T]\to {\mathbb R}^d\,:\, \phi(t) = x_0 + \int_0^t \dot\phi(s)\,ds,\,\, \int_0^T |\dot \phi(t)|^2\,dt < \infty \right\}
\end{equation*}
\end{theorem}
\vskip10pt

\begin{proof}
By Theorem~\ref{theorem1} and assumptions (ii)-(iii),
\begin{equation}\label{aa}
{\mathbb E}\left[\frac{d\mathbb P}{d\tilde {\mathbb P}}\,1_{{\mathcal A}_\epsilon}\right] 
\le e^{\epsilon^{-1}\left(V(x_0) - {\tilde V}(x_0)\right) + TM}{\mathbb P}({\mathcal A}_\epsilon)
\end{equation}
Using assumption (i), choose $\epsilon > 0$ so that
\begin{equation*}
{\tilde V}(x_0)-V(x_0) \ge \epsilon TM
\end{equation*}
Then from \eqref{var1}, \eqref{var2} and \eqref{reduct}, 
\begin{equation}\label{bb}
 \frac{{\tilde {\hbox{Var}}}({\tilde \Theta}_\epsilon)}{\hbox{Var}(\Theta_\epsilon)} = \frac{{\tilde{\hbox{Var}}} \left(\frac{{d\mathbb P}}{d{\tilde {\mathbb P}}}\,1_{{\mathcal A}_\epsilon}\right)}{{\hbox{Var}}(1_{{\mathcal A}_\epsilon})} 
\le \frac{{\mathbb E}\left[\frac{d\mathbb P}{d\tilde {\mathbb P}}\,1_{{\mathcal A}_\epsilon}\right]}{{\mathbb P}({\mathcal A}_\epsilon)}
\end{equation}
Comparing with \eqref{aa} with \eqref{bb}, 
\begin{equation*}
\frac{{\tilde {\hbox{Var}}}({\tilde \Theta}_\epsilon)}{\hbox{Var}(\Theta_\epsilon)}
\le e^{\epsilon^{-1}\left(V(x_0) - {\tilde V}(x_0)\right) + TM}  
\end{equation*}
From Definition~\ref{definition1} and continuity 
of $\nabla V$ it follows that ${\mathcal A}_\epsilon$ is a continuity 
set \cite{Ofer} with respect to the rate function
\begin{equation*}
 \phi \to \begin{cases} \frac{1}{2}\int_0^T \left|\dot \phi(t)+\nabla V(\phi(t))\right|^2 dt, & \phi \in H_{x_0, T}^1 \\
           \infty, & \phi \notin H_{x_0, T}^1
          \end{cases}
\end{equation*}
Therefore 
\begin{equation*}
 \lim_{\epsilon \to 0} \epsilon \log {\mathbb P}({{\mathcal A}_\epsilon}) = -I_V(x_0)
\end{equation*}
A simple calculation now leads to \eqref{asympopt}.
\end{proof}
\vskip10pt

\begin{figure}
\begin{center}
\includegraphics[scale=0.4]{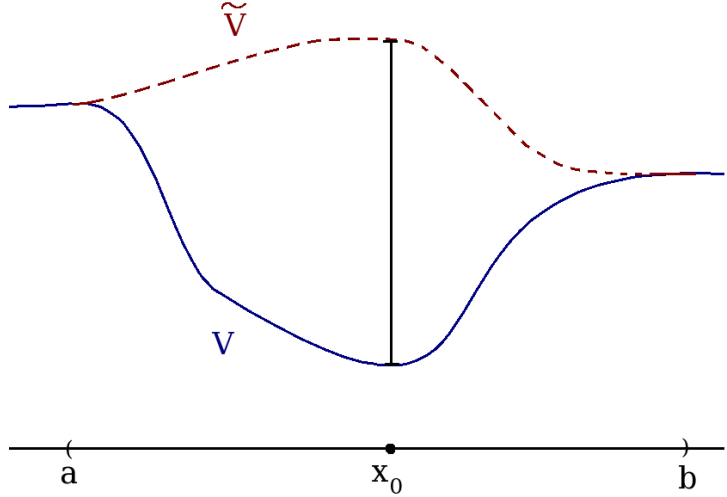}
\end{center}
\caption{Using the sampling potential $\tilde V$ 
to estimate the rare observable ${\mathbb P}({X_T^\epsilon \notin D})$, for $D = (a,b)$. 
The logarithm of the reduction of variance compared to standard Monte Carlo 
is closely related to the length of the vertical line.}
\end{figure}

Theorem~\ref{theorem3} shows that by choosing an sampling potential ${\tilde V}$ which reduces 
the depth of the potential well around $x_0$ and which agrees with $V$ outside the well, 
the probability that the process \eqref{smallnoise1} is outside the well at time $T$ can be estimated 
with an exponentially reduced variance compared to standard Monte Carlo. The variance is 
reduced by a factor proportional to 
\begin{equation*}
 \exp\left[\epsilon^{-1}(V(x_0)-{\tilde V}(x_0))\right]
\end{equation*}
See Figure 1.

In Theorem~\ref{theorem4} below is shown that if the well has a flat boundary, then an 
asymptotically optimal scheme is obtained 
by inverting the potential well inside $D$;
see Figure 2.
\vskip10pt

\begin{theorem}
 \label{theorem4}
Assume $V \in C_b^2({\mathbb R}^d)$ and that $\nabla V(x) \equiv 0$ on $\partial D$. Then 
WLOG we may take $V(x) \equiv 0$ on $\partial D$. 
Define 
\begin{equation*}
 {\tilde V}(x) = \begin{cases} -V(x) & \hbox{ if } x \in D \\ V(x) & \hbox{ if }x \notin D \end{cases}
\end{equation*}
and assume ${\tilde V} \in C^2({\mathbb R}^d)$. 
Define 
\begin{equation*}
K \equiv  \sup_{x\in D} |\Delta {\tilde V}| = \sup_{x\in D} |\Delta V| 
\end{equation*}
Let $\tilde {\mathbb P}$ be the reference measure under which $X_t^\epsilon$ satisfies \eqref{smallnoise2}, 
and define ${{\mathbb P}}$ as in \eqref{chg1}, so that under 
${{\mathbb P}}$, $X_t^\epsilon$ satisfies \eqref{smallnoise1}. \textup{(}The dependence of 
$\mathbb P$ and $\tilde {\mathbb P}$ on $\epsilon$ is suppressed.\textup{)} 
Furthermore assume the solution $y = y(t)$ to the IVP
\begin{equation}\label{assumpt}
 \frac{dy}{dt} = -\nabla {\tilde V}(y),\hskip10pt y(0) = x_0
\end{equation}
satisfies $y(T) \notin D$. Then 
 \begin{equation*}
\lim_{\epsilon\to 0} \epsilon \log \Lambda_\epsilon
= 0
\end{equation*}
\end{theorem}
\vskip10pt

\begin{proof}
From the proof of Theorem~\ref{theorem3}, 
\begin{eqnarray}\label{bound1}
 \lim_{\epsilon\to 0} \epsilon \log \Lambda_\epsilon
&\le& 2V(x_0) + I_V(x_0) \\ \label{bound2}
&=& 2V(x_0) - \lim_{\epsilon \to 0}\epsilon \log {\mathbb P}({\mathcal A}_\epsilon)
\end{eqnarray}
Now Theorem~\ref{theorem1} gives 
\begin{equation}\label{here}
 {\mathbb P}({\mathcal A}_\epsilon) \ge e^{2\epsilon^{-1}V(x_0)-T{K}}{\tilde {\mathbb P}}({\mathcal A}_\epsilon)
\end{equation}
From Definition~\ref{definition1} and continuity 
of $\nabla {\tilde V}$ it follows that ${\mathcal A}_\epsilon$ is a continuity 
set with respect to the rate function
 \begin{equation*}
 \phi \to \begin{cases} \frac{1}{2}\int_0^T \left|\dot \phi(t)+\nabla {\tilde V}(\phi(t))\right|^2 dt, & \phi \in H_{x_0, T}^1 \\
           \infty, & \phi \notin H_{x_0, T}^1
          \end{cases}
\end{equation*}
Therefore by assumption \eqref{assumpt},  
\begin{equation*}
\lim_{\epsilon \to 0} \epsilon \log {\tilde {\mathbb P}}({\mathcal A}_\epsilon) = -I_{\tilde V}(x_0) = 0
\end{equation*}
From \eqref{here} it follows that 
\begin{equation}\label{this}
 \lim_{\epsilon \to 0} \epsilon \log  {\mathbb P}({\mathcal A}_\epsilon) \ge 2V(x_0)
\end{equation}
By comparing \eqref{this} with \eqref{bound1}-\eqref{bound2}, 
\begin{equation*}
\lim_{\epsilon\to 0} \epsilon \log \Lambda_\epsilon \le 0
\end{equation*}
From Jensen's inequality $\Lambda_\epsilon \ge 1$ and the result follows.
\end{proof}
\vskip10pt

\begin{figure}
\begin{center}
\includegraphics[scale=0.4]{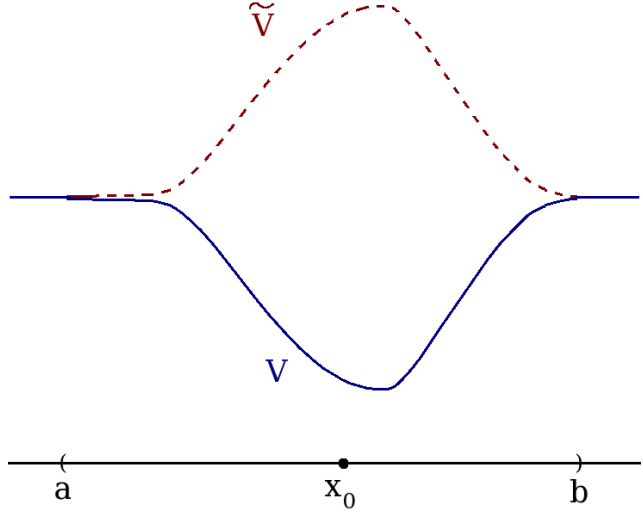}
\end{center}
\caption{Using the sampling potential $\tilde V$ 
to estimate the rare observable ${\mathbb P}({X_T^\epsilon \notin D})$, for $D = (a,b)$. 
If $X_T^0 \notin D$ (that is, if the process lands outside of $D$ with deterministic dynamics) 
then the reduction in variance is asymptotically optimal.}
\end{figure}

Although the assumptions of Theorem~\ref{theorem4} 
are very restrictive, the result suggests what changes in potential should be most effective 
more generally. 
In particular, by inspecting \eqref{chg1} and the proof of Theorem~\ref{theorem3}, one sees that in 
choosing an optimal $\tilde V$, there is a competition between maximizing ${\tilde V}(x_0)$ 
while also minimizing $|\nabla \tilde V(x)|$ and maximizing $\Delta {\tilde V}(x)$ for $x \in D$, 
in the sense that when any two of these three is fixed, optimizing the third reduces the variance. 
Here it is assumed that $\tilde V$ agrees with $V$ outside $D$. Theorem~\ref{theorem3} shows that in 
the small noise limit, maximizing $\Delta {\tilde V}(x)$ becomes  
irrelevent; Theorem~\ref{theorem4} suggests that it may be near optimal to choose a $\tilde V$ 
which maximizes ${\tilde V}(x_0)$ while also minimizing $||\nabla{\tilde V}(x)| -  |\nabla V(x)||$ 
for $x\in D$.

\section{Example}\label{ex}

Consider the one-dimensional overdamped Langevin SDE 
\begin{equation}\label{oned}
 dX_t = -\frac{d}{dx}V(X_t)\,dt + dW_t, \hskip10pt X_0 = 0, \hskip10pt 0\le t \le T
\end{equation}
where $V(x) = -\cos x-1$ and $W_t$ is a Wiener process.
Let $T=1$, define 
\begin{equation}
 {\mathcal A} = \{X_\cdot \, :\, X_T \notin (-\pi,\pi)\}
\end{equation} 
and suppose the probability of interest is ${\mathbb P}({\mathcal A})$ where 
${\mathbb P}$ is the probability of the process \eqref{oned}. Consider the 
importance sampling scheme of Section~\ref{imp}, with sampling potentials  
\begin{align*}
&{\tilde V}^A(x) = \begin{cases}
0, & \hbox{if }x \in (-\pi,\pi) \\
V(x), & \hbox{otherwise }
\end{cases} \\
&{\tilde V}^B(x) = \begin{cases}
-V(x), & \hbox{if }x \in (-\pi,\pi) \\
V(x), & \hbox{otherwise }
\end{cases}
\end{align*}
Let ${\tilde {\mathbb P}}^A$ be the reference measure under which $X_t$ satisfies 
\begin{equation}\label{oned1}
 dX_t = -\frac{d}{dx}{\tilde V}^A(X_t)\,dt + d{\tilde W}_t^A, \hskip10pt X_0 = x_0, \hskip10pt 0\le t \le T
\end{equation}
and let ${\tilde {\mathbb P}}^B$ be the reference measure under which $X_t$ satisfies 
\begin{equation}\label{oned2}
 dX_t = -\frac{d}{dx}{\tilde V}^B(X_t)\,dt + d{\tilde W}_t^B, \hskip10pt X_0 = x_0, \hskip10pt 0\le t \le T
\end{equation}
where $d{\tilde W}_t^A$ and $d{\tilde W}_t^B$ are ${\tilde {\mathbb P}}^A$- and 
${\tilde {\mathbb P}}^B$-Wiener processes, respectively. 
Then under ${\mathbb P}$ defined by \eqref{chg1}, $X_t$ satisfies \eqref{oned}. 
The following table compares standard Monte Carlo estimates of ${\mathbb P}({\mathcal A})$, 
using $\Theta$ defined in \eqref{theta1}, to estimates 
from the scheme outlined in Section~\ref{imp}. The importance sampling estimators ${\tilde \Theta}^A$ and 
${\tilde \Theta}^B$ corresponding to ${\tilde {\mathbb P}}^A$ and ${\tilde {\mathbb P}}^B$ are defined as in \eqref{theta2}.
Samples are obtained using 
Euler approximations of \eqref{oned} and \eqref{oned1}-\eqref{oned2} with step size $h = 10^{-5}$  
and Riemann approximations 
of \eqref{chg1} with mesh size $\tau$.  

\vskip15pt 

\begin{tabular}{ l | c| c | c | c | l}
  estimator &  potential & $N$ (\# of samples) & $\tau$ & sample average & sample variance \\ \hline 
  $\Theta$ & $V$ & $10^8$ & N/A & 0.000192 & 0.000192 \\
  ${\tilde \Theta}^A$ & ${\tilde V}^A$ & $10^7$ & $10^{-1}$ & 0.000195 & 0.000022 \\
  ${\tilde \Theta}^A$ & ${\tilde V}^A$ & $10^7$ & $10^{-2}$ & 0.000193 & 0.000022 \\
  ${\tilde \Theta}^A$ & ${\tilde V}^A$ & $10^7$ & $10^{-3}$ & 0.000193 & 0.000022 \\
  ${\tilde \Theta}^B$ & ${\tilde V}^B$ & $10^7$ & $10^{-1}$ & 0.000204 & 0.0000039 \\
  ${\tilde \Theta}^B$ & ${\tilde V}^B$ & $10^7$ & $10^{-2}$ & 0.000196 & 0.0000039 \\
  ${\tilde \Theta}^B$ & ${\tilde V}^B$ & $10^7$ & $10^{-3}$ & 0.000197 & 0.0000039 \\
\end{tabular}
\vskip15pt

\vskip15pt

\noindent Note that sampling with either ${\tilde V}^A$ or ${\tilde V}^B$ reduces the variance significantly. 
The greater reduction in variance is obtained by sampling with ${\tilde V}^B$. This is 
consistent with Theorem~\ref{theorem4}, which suggests ${\tilde V}^B$ is 
asymptotically optimal. Note that ${\tilde V}^A$ and ${\tilde V}^B$ do not quite satisfy the conditions of the 
theorem since 
\begin{equation*}
 \frac{\partial^2}{\partial x^2}{\tilde V}^A(\pm \pi), \hskip5pt \frac{\partial^2}{\partial x^2}{\tilde V}^B(\pm \pi)
\end{equation*}
do not exist, yet the scheme is nonetheless accurate and effective. One 
suspects the assumption ${\tilde V} \in C_b^2({\mathbb R}^d)$ in Theorem~\ref{theorem3} and 
Theorem~\ref{theorem4} can be relaxed to ${\tilde V} \in C_b^1({\mathbb R}^d)$ on $\partial D$;  
this generalization is not pursued here. Though $\epsilon = 1$ here is 
``far'' from zero, one expects that $P({\mathcal A})$ being small means exactly 
that one is effectively in the small noise regime.

\section{Conclusion}
The problem of estimating small probabilities of the overdamped Langevin process, a well-known and 
important model of physical systems, is explored. Since standard Monte Carlo techniques are often impractical in this setting, it is useful to 
have alternative means of estimating averages of observables, in particular of transition probabilities. 
This paper explores small transition probabilities of Langevin processes in two asymptotic regimes: 
$t\approx 0$ and $\beta =2\epsilon^{-1} \approx \infty$. A first-order accurate asymptotic correction to 
transition probability densities as $t\to 0$ is proved in Theorem~\ref{theorem2}, and an importance sampling 
technique for estimating escape probabilities as $\beta\to \infty$ is shown in Theorem~\ref{theorem3} to perform 
exponentially better than standard Monte Carlo. It is shown that this 
technique is asymptotically optimal in some cases (Theorem~\ref{theorem4}). The importance sampling scheme 
has the virtue of requiring nearly neglibible added computation (compared with standard Monte Carlo) during simulations, 
and it is shown to be effective in a simple numerical example.


\begin{thebibliography}{17}

\bibitem{Oksendal} B.K. {\O}ksendal, Stochastic Differential Equations: An Introduction with Applications, Sixth Edition, Springer (2003)

\bibitem{Lelievre} T. Leli{\`e}vre, M. Rousset and G. Stoltz, Free energy computations: a mathematical perspective, Imperial College Press (2010)

\bibitem{Nelson} E. Nelson, Dynamical Theories of Brownian Motion, Princeton University Press, Second Edition (1967)


\bibitem{Freidlin} M.I. Freidlin and A.D. Wentzell, Random Perturbations of Dynamical Systems, Springer, (1984)

\bibitem{Ofer} A. Dembo and O. Zeitouni, Large Deviations Techniques and Applications, Springer, Second Edition (1998)

\bibitem{Dupuis1} P. Dupuis and H. Wang, Subsolutions of an Isaacs equation and efficient schemes for importance sampling, 
Mathematics of Operations Research, Vol. 32, No. 3, 732-759 (2007)

\bibitem{Dupuis} P. Dupuis, K. Spiliopoulos, and H. Wang, Importance Sampling for Multiscale Diffusions, arXiv:1107.5448v1

\bibitem{Weare} E. Vanden-Eijnden and J. Weare, Rare event simulation with vanishing error for small noise diffusions, Preprint

\bibitem{Guasoni} P. Guasoni and S. Robertson, Optimal importance sampling with explicit formulas in continuous time, Finance and Stochastics 12, 1-19 (2008)

\bibitem{Koralov} L. Koralov and Y. Sinai, Theory of Probability and Random Processes, Springer, Second Edition (2007)

\bibitem{Grigoriu} M. Grigoriu, Stochastic Calculus: Applications in Science and Engineering, Birkh\"{a}user (2002)


\bibitem{Girsanov} I.V. Girsanov, On transforming a certain class of stochastic processes by absolutely continuous substitution of measures, Theory of Probability and its Applications, Volume V, No. 3 (1960)


\bibitem{Wang} L. Wang and K. P\"{o}tzelberger, Boundary Crossing Probability for Brownian Motion and General Boundaries


\bibitem{Siegmund} D. Siegmund, Boundary crossing probabilities and statistical applications. Ann. Statist.  14,  361-404 (1986) 

\bibitem{Baldi} Asymptotics of Hitting Probabilities for General One-Dimensional Pinned Diffusions, Annals of Applied Probability, Vol 12, No 3, 1071-1095 (2002)

\bibitem{Asmussen} S. Asmussen, P.W. Glynn, Stochastic Simulation: Algorithms and Analysis, Stochastic Modelling and Applied Probability, Springer Verlag, Chapter VI (2007)

\bibitem{Bucklew}  J.A. Bucklew, Introduction to Rare Event Simulation, Springer (2004)




\end{thebibliography}
\end{document}